\documentclass[pre,showpacs,preprint,superscriptaddress]{revtex4-1}
\usepackage{amssymb,graphicx,amsbsy,amsmath}
\usepackage[usenames]{color}

\begin{document}
\title{Residual discrete symmetry of the five-state clock model}
\author{Seung Ki Baek}
\email[E-mail: ]{seungki@kias.re.kr}
\affiliation{School of Physics, Korea Institute for Advanced Study, Seoul
130-722, Korea}
\author{Harri M\"akel\"a}
\affiliation{QCD Labs, COMP Centre of Excellence, Department of Applied Physics, Aalto University, P.O. Box 13500, FI-00076 AALTO, Finland}
\author{Petter Minnhagen}
\affiliation{Department of Physics, Ume{\aa} University, 901 87 Ume{\aa},
Sweden}
\author{Beom Jun Kim}
\email[E-mail: ]{beomjun@skku.edu}
\affiliation{BK21 Physics Research Division and Department of Physics, Sungkyunkwan University, Suwon 440-746, Korea}

\begin{abstract}
It is well-known that the $q$-state clock model can exhibit a
Kosterlitz-Thouless (KT) transition if $q$ is equal to or greater
than a certain threshold, which has been believed to be five.
However, recent numerical studies indicate that helicity modulus does not
vanish in the high-temperature phase of the five-state clock model as
predicted by the KT scenario.
By performing Monte Carlo calculations under the
fluctuating twist boundary condition, we show that it is because the
five-state clock model does not have the fully continuous $U(1)$ symmetry
even in the high-temperature phase while the six-state clock model does.
We suggest that the upper transition of the five-state clock model is
actually a weaker cousin of the KT transition so that it is $q \ge 6$ that
exhibits the genuine KT behavior.
\end{abstract}


\maketitle

\section{Introduction}
\label{sec:intro}

Theories of critical phenomena in low dimensions are now quite well
established due to the concepts of the renormalization group (RG) and
universality. The two-dimensional (2D) Ising model and the 2D $XY$ model are
two classical model systems that have considerably enhanced our
understanding of these concepts. While the former exhibits an order-disorder
phase transition, the physics of the latter model is better described in
terms of vortex-antivortex unbinding, where the
Kosterlitz-Thouless (KT) picture emerges~\cite{kt,*kost}.
Closely related to these models, one may also consider
the 2D $q$-state clock model defined by the following Hamiltonian:
\begin{equation}
H = - J \sum_{\left< ij \right>} \cos(\theta_i - \theta_j),
\label{eq:h}
\end{equation}
where $J$ means interaction strength, the summation is over the nearest
neighbors, and $\theta_i = 2\pi n_i/q$ with $n_i = 0, \ldots, q-1$.
As can be easily seen, this model has been
extensively studied as a bridge between the Ising model ($q=2$) and the $XY$
model ($q \rightarrow \infty$). But it does not mean that $q \rightarrow
\infty$ is actually required for a KT transition to be observed:
Rather, the KT behavior becomes possible for $q > q_c$ with an
intermediate quasi-ordered massless phase between an ordered phase at low
temperature and a disordered phase at high temperature. A hand-waving
RG argument is that domain walls get so floppy to
effectively provide block spins with the continuous $U(1)$ symmetry over
large length scales~\cite{einhorn}.
In order to estimate $q_c$, let us generalize Eq.~(\ref{eq:h}) so that
\[ H = \sum_{\left< ij \right>} V(\theta_i - \theta_j), \]
where the spin-interaction potential $V$ has the $Z_q$ symmetry.
The Villain $q$-state clock model chooses, for example,
\[ V(\phi) = -\frac{J}{\beta} \ln \left\{ \sum_{n=-\infty}^{\infty}
\exp\left[ -\beta (\phi - 2\pi n)^2/2 \right] \right\},\]
where  $\beta \equiv 1/(k_B T)$ with the Boltzmann constant $k_B$ and
temperature $T$. This potential form has been introduced to separate the
vortex degrees of freedom from the spin-wave degrees of freedom as an 
approximate version of the $XY$ model~\cite{savit}.
This Villain clock model on the square lattice possesses
self-duality~\cite{savit}, predicting that $q_c = 4$ based on the following
argument~\cite{elit}: Suppose that there are only two phases. Then, the
phase transition should occur at the dual point $T^{\rm Villain}_d =
2\pi/q$. The Villain $XY$ model has a transition temperature around $T^{\rm
Villain}_{\rm KT} \sim 1.35$. For $q > 4$, therefore, $T^{\rm Villain}_d$
becomes lower than $T^{\rm Villain}_{\rm KT}$. However, this is against the
correlation inequality~\cite{savit}, which says that
\[ \left< \cos(\theta_i - \theta_j) \right>^{\rm Villain}_q \ge \left<
\cos(\theta_i - \theta_j) \right>^{\rm Villain}_{XY}, \]
where $i$ and $j$ are arbitrary spin indices and the brakets mean
statistical averages of the Villain $q$-state clock model and that of the
Villain $XY$ model, respectively, at the same temperature.
It expresses an intuitively obvious idea that ordering is
generally weaker in the latter model since it has a greater number of
possible spin configurations, and this sort of statement holds even if we
work with the cosine potential. In short, the transition temperature of the
Villain $q$-state clock model cannot be lower than $T^{\rm Villain}_{\rm
KT}$, contradicting the duality argument in the two-phase picture.
Therefore, the two-phase picture is not valid and we are forced to have
three phases at least for $q > 4$. The same conclusion can be drawn for
the triangular and honeycomb lattices as well~\cite{cardy80}.

\begin{figure}
\includegraphics[width=0.45\textwidth]{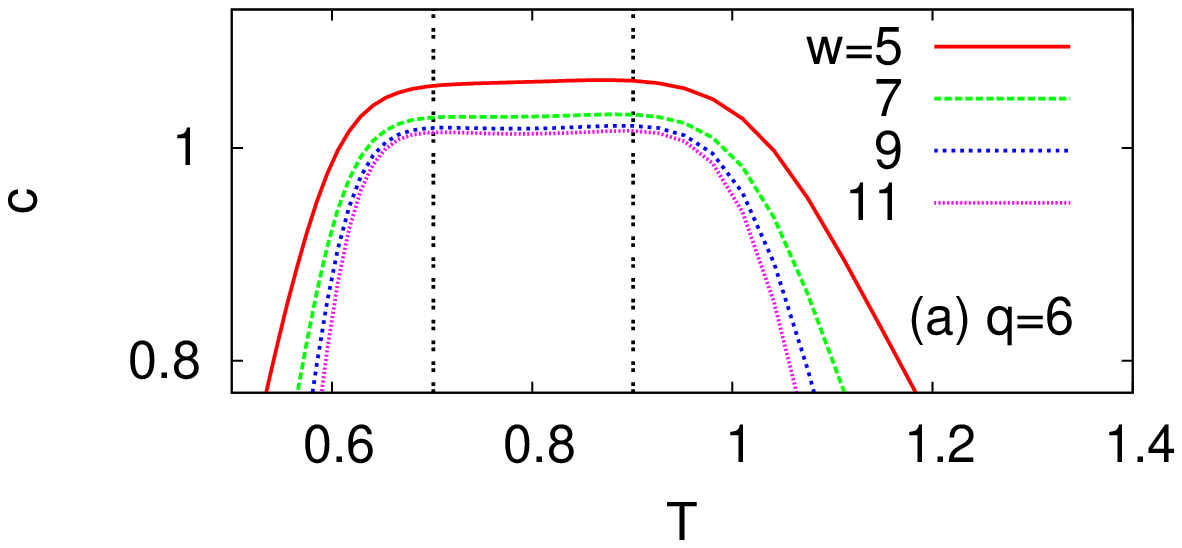}
\includegraphics[width=0.45\textwidth]{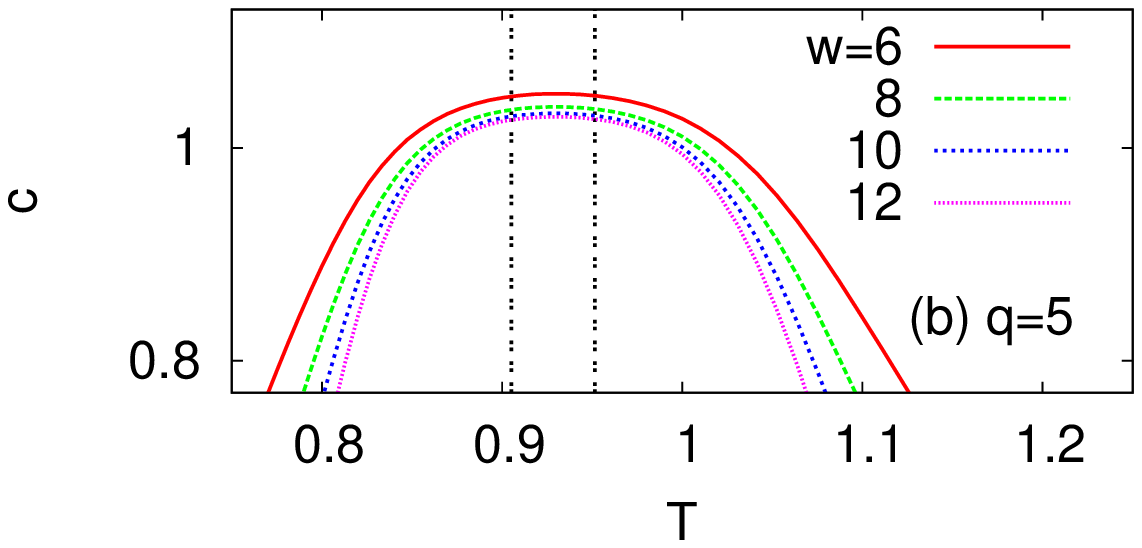}
\caption{(Color online) (a) Conformal charge of the six-state clock model
estimated by using semi-infinite strips with different widths $w$.
The vertical lines represent $T_{c1}$ and $T_{c2}$ obtained in
Ref.~\cite{tomita}. (b) The same quantity for the five-state clock model,
where the vertical lines represent $T_{c1}$ and $T_{c2}$ obtained in
Ref.~\cite{papa1}.
The free energy of each finite-sized system is calculated by using the
transfer-matrix method~\cite{blote1,*blote2,*foster} and the bulk free
energy is obtained by the Bulirsch-Stoer extrapolation
algorithm~\cite{henkel88}. Note the convergence to $c=1$ as $w$ grows in
both the cases.}
\label{fig:transfer}
\end{figure}

Since this argument relies on the specific choice of $V(\phi)$ in the
Villain form, one
might wonder if the same conclusion can be drawn for the usual clock model
with the cosine potential. For instance, a recent claim was $q_c =
7$ based on helicity modulus~\cite{lapilli}, though more recent calculations
have confirmed a normal KT transition for $q = 6$~\cite{comment}. Since
the four-state clock model exactly belongs to the Ising universality
class~\cite{suzuki}, the only remaining case to consider is $q=5$. Although
the Migdal-Kadanoff RG transformation predicts
only two phases~\cite{nish}, there is ample evidence that this five-state
clock model does have three phases~\cite{tobo,bonnier3,papa1}, and
theoretical arguments for the three-phase picture have been motivated and
supported by these observations. We first calculate the conformal charge $c$
from the leading finite-size correction of the free energy on semi-infinite
strips of the square geometry~\cite{cardy86}, and this also indicates the
existence of a temperature region with $c=1$ as in the six-state clock model
[compare Fig.~\ref{fig:transfer}(a) and \ref{fig:transfer}(b)].
Even if not completely rigorous, a convincing argument for the
three-phase picture originates from a field-theoretic
construction~\cite{fateev}, which shows that there exists a conformally
invariant point with conformal charge $c=8/7$ between the Potts model and
the clock model, and conjectures that this is where the two critical lines
merge~\cite{*[{This special point corresponds to $p=2.01677$ and $\beta =
5.27413$ in terms of the generalized clock model parameterized by $p$ as
found in }] [{.}] q8}.
This conjecture is found consistent with numerical
results~\cite{alcaraz,bonnier1,bonnier2}.
In short, the dominant consensus is that the
five-state clock model belongs to the same universality class as the
Villain five-state clock model, where one KT transition and its dual separate
three phases.

\begin{figure}
\includegraphics[width=0.45\textwidth]{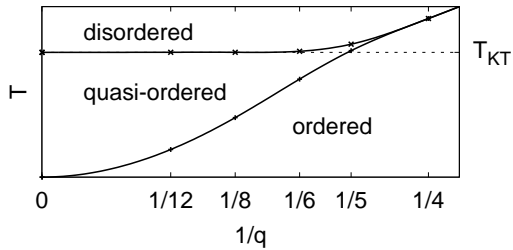}
\caption{Phase diagram of the $q$-state clock model. The transition points
for $q>4$ are taken from Refs.~\cite{papa1,tomita,hasen1}, and the solid
lines are guides to eyes. Note that both the transition points for $q=5$ are
above the dotted horizontal line, representing $T_{\rm KT}$.}
\label{fig:phase}
\end{figure}

However, we have observed that helicity modulus of the five-state clock
model does not vanish at the upper transition~\cite{clock},
which casts doubt on its KT nature.
That is, if the system underwent a KT transition, this quantity would show a
jump to zero~\cite{nelson,*minnhagen}.
Here, the helicity modulus is defined as response to an infinitesimal
twist $\Delta$ across the system in the $x$ direction:
\begin{equation}
\Upsilon \equiv \left. \frac{\partial^2 F}{\partial \Delta^2}
\right|_{\Delta = 0} = \left< e \right> - L^2\beta \left<s^2\right>,
\label{eq:y}
\end{equation}
where $e \equiv L^{-2} \sum_{\left<ij\right>_x} V''(\theta_i - \theta_j)$
and $s \equiv L^{-2} \sum_{\left<ij\right>_x} V'(\theta_i - \theta_j)$ are
summed over all the links in the $x$ direction.
Although extensive Monte Carlo
calculations have shown that all the other features of the transition are
not inconsistent with the KT picture~\cite{papa1,papa2}, these
calculations raise further questions since
the transition temperatures are estimated as $T_{c1} = 0.9051(9)$ and
$T_{c2} = 0.9515(9)$, respectively: Estimates of the
transition temperature $T_{\rm KT}$ of the $XY$ model vary among
authors~\cite{olsson,tomita,hasen1,*hasen2}, but one can safely say $T_{\rm
KT} < 0.894$. The point is that $T_{\rm KT}$ is still well below the
lower transition temperature of the five-state clock model, as well as
the dual point $T_d \approx 0.9291$~\cite{nish,wu} (see Fig.~\ref{fig:phase}).
This implies a sort of \emph{prematurity} in the emergence of three phases
in the five-state clock model in that the duality cannot enforce it as in
the Villain clock model.
In addition, when the mass gap in the Hamiltonian formulation has a
singularity of the following form:
\[ m(\tau) \sim \exp(C \tau^{-\sigma}), \]
where $\tau$ is the reduced temperature and $C$ is a constant, the KT
prediction of $\sigma = 1/2$~\cite{kt,*kost} appears consistent only when
$q \ge 6$, whereas it converges to $\sigma \approx 0.2$
at $q=5$~\cite{elit,bonnier1}. Note that a
second-order phase transition would be represented by $\sigma = 0$.
All these signatures suggest something inherently different in the
five-state clock model from the six-state clock model as well as from the
four-state clock model.
To sum up, the situation is that
the three-phase picture looks convincing in the five-state clock model,
but that consistency with the KT picture is still doubtful.

In this work, we explicitly show that the nonvanishing helicity modulus of
the five-state clock model
means a residual five-fold symmetry, which is not completely washed away
by spin fluctuations even at $T \ge T_{c2}$. It implies a possibility that
the continuous $U(1)$ symmetry as in the $XY$ model sets in only when $T
\rightarrow \infty$, which means a difference from the Villain
formulation. After presenting our numerical results in Sec.~\ref{sec:ftbc},
we discuss a theoretical attempt to explain this feature by introducing
the second length scale of vortex composites in Sec.~\ref{sec:discuss}, and
then conclude this work.

\section{Fluctuating twist boundary condition}
\label{sec:ftbc}

In the fluctuating twist boundary condition (FTBC)~\cite{ftbc,*ftbc2},
there exists
interaction among the boundary spins as in the periodic boundary condition.
A difference is that here we apply a gauge field that adds twist $\Delta$ in
the interaction potential. If $\Delta$ is fixed at zero, it is identical to
the periodic boundary condition, whereas if it is fixed at $\pi$, it
corresponds to the anti-periodic boundary condition. As the name indicates,
the FTBC lets the twist $\Delta$ also fluctuate in time following the
Metropolis update rule. The free energy is written as $F = F(T, \Delta)$,
and the probability to observe $\Delta$ at a given $K$ will be $P(T,\Delta)
\propto \exp[-\beta F(T,\Delta)]$~\cite{bennett}. In this way, we measure
how the free energy changes as $\Delta$ varies. When $\Delta$ is a
continuous variable, since $F(K,\Delta) = F(K,0) + \frac{\Delta^2}{2}\left.
\frac{\partial^2 F}{\partial \Delta^2} \right|_{\Delta = 0} + \ldots$, this
method provides an estimate of the helicity modulus defined in
Eq.~(\ref{eq:y}).

\begin{figure}
\includegraphics[width=0.45\textwidth]{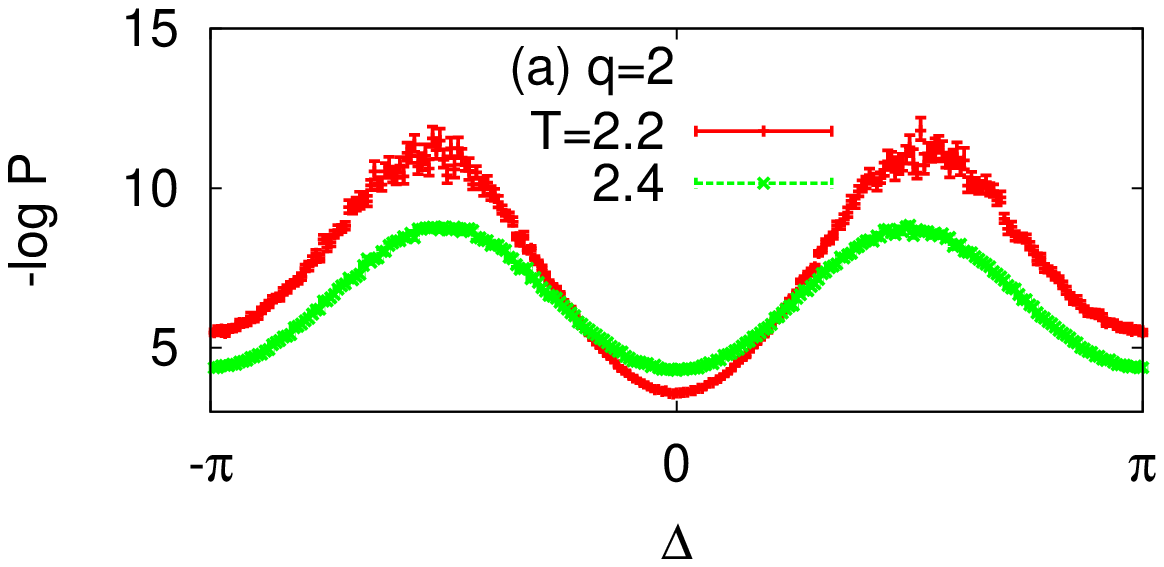}
\includegraphics[width=0.45\textwidth]{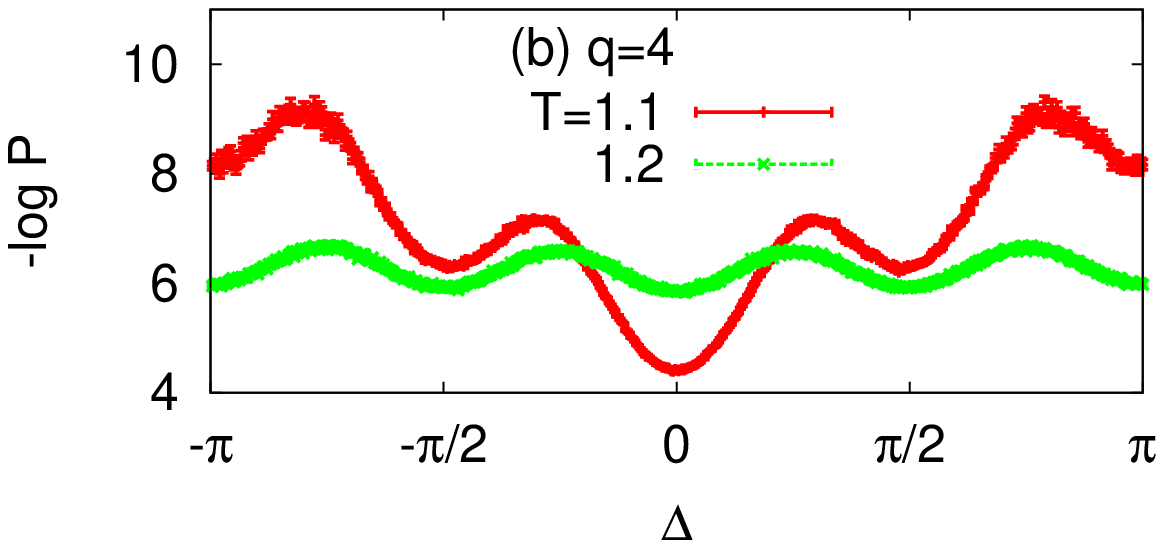}
\includegraphics[width=0.45\textwidth]{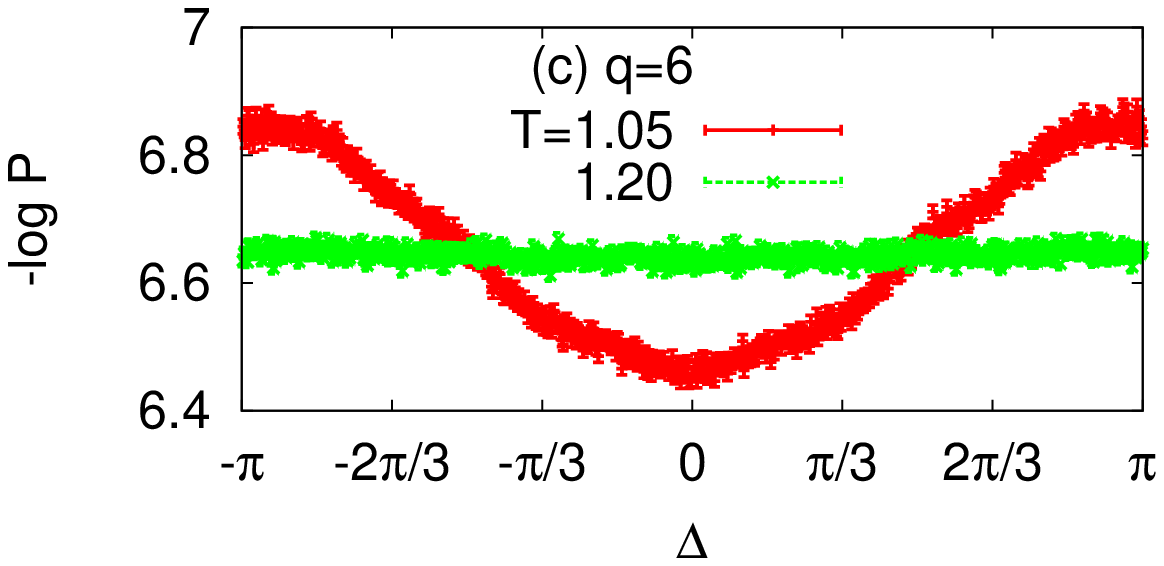}
\includegraphics[width=0.45\textwidth]{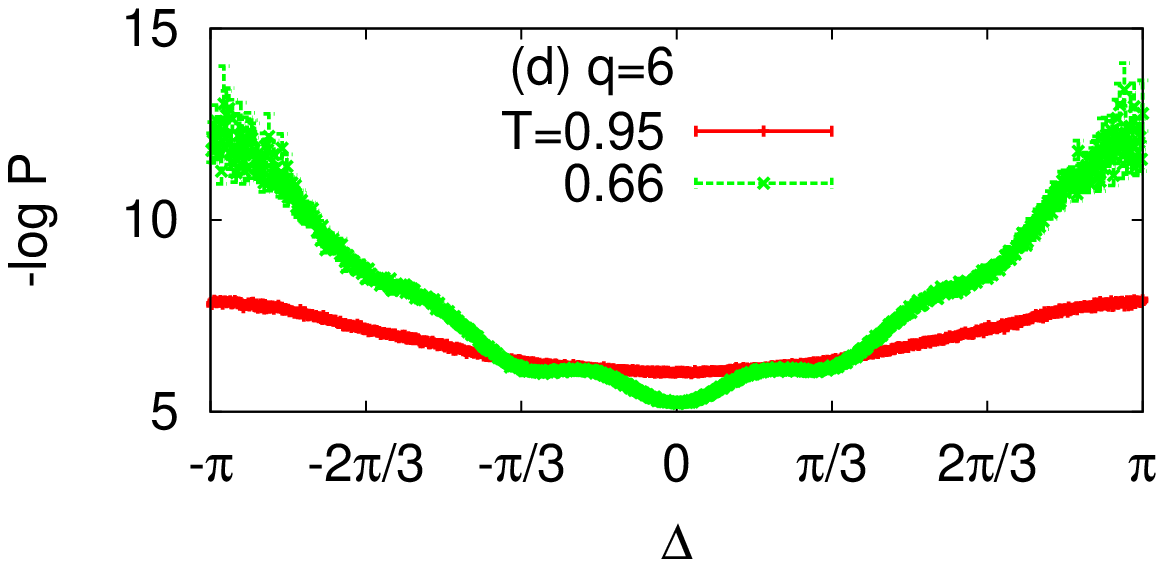}
\caption{(Color online) Free energy $F(T,\Delta) \propto -\ln P(T,\Delta)$
where $P(T,\Delta)$ is probability of observing twist $\Delta$ under the
fluctuating twist boundary condition at temperature $T$. In practice, the
twist $\Delta$ is given as a $(128\times q)$-state clock spin to simulate a
continuous variable.
(a) The discrete symmetry is clearly detected in
the disordered phase of the Ising model, and (b) the same statement can be
made for the four-state clock model. (c) On the other hand, the discrete
symmetry changes to the continuous $U(1)$ symmetry in the six-state
clock model and (d) manifests itself only at much lower $T$.
The system size is taken as $L=32$ in every case.}
\label{fig:ftbc}
\end{figure}

Let us begin with the 2D Ising model ($q=2$) on the square lattice.
Since $\left<s^2\right>$ is identically zero, its helicity modulus
$\Upsilon = \left< e \right>$ is proportional to the internal energy
density, which obviously does not vanish at any temperature [see
Eq.~(\ref{eq:y})]. On the other hand, the
free-energy difference between the periodic boundary condition and the
anti-periodic boundary condition becomes exponentially small as the system
size grows if the temperature is higher than
$T^{\rm Ising}_c = 2/\ln(1+\sqrt{2}) \approx 2.269$ in units of $J/k_B$.
These two facts immediately suggest that
the free energy $F(\Delta)$ has a smooth barrier between $\Delta =
0$ and $\pi$ and another between $\Delta = 0$ and $-\pi$ by symmetry.
This speculation is directly confirmed in Fig.~\ref{fig:ftbc}(a) obtained by
using the FTBC, where we see that the
system is disordered but preserves its up-down symmetry. It is also
intuitively clear that the system begins to prefer $\Delta = 0$ to $\Delta =
\pm\pi$ as $T$ passes $T^{\rm Ising}_c$ from above so that $F(\Delta)$ changes
its shape as shown in Fig.~\ref{fig:ftbc}(a).
To understand the nonvanishing $\Upsilon$ on physical grounds,
it is instructive to recall that the free energy is written as $F = U-TS$
with internal energy $U=U(T,\Delta)$ and entropy $S=S(T,\Delta)$. Since the
first term in Eq.~(\ref{eq:y}) is related to the
second derivative of $U$, the second term describing fluctuations should be
related to the entropic contribution to the free-energy change. We can
say that if $\Delta$
is very small, it is difficult for the system to excite domain walls whose
entropy could compensate for the energy change.
In this sense, the nonvanishing $\Upsilon$ means that enough domain walls, a
discrete version of spin waves, are not generated to cancel out
the external perturbation $\Delta$, possibly due to the finite energy scale needed
to excite them.

The four-state clock model with interaction strength $J$ has the same
partition function as two independent Ising systems with interaction
strength $J/2$~\cite{suzuki}.
This model has $c=1$ since each Ising model carries conformal charge
$c=1/2$~\cite{boyan}, but the massless phase exists only at a single
point so the model has only two phases separated
at $T^{q=4}_c = T^{\rm Ising}_c/2 \approx 1.135$. It is also
obvious from this equivalence that $\Upsilon$ does not vanish at $T >
T_c$, and we see how the system preserves its four-fold symmetry in
Fig.~\ref{fig:ftbc}(b). In terms of vortices, this is related to the fact
that a vortex is not a really independent object but merely results from two
independent Ising domain walls that happen to cross at a point.

\begin{figure}
\includegraphics[width=0.45\textwidth]{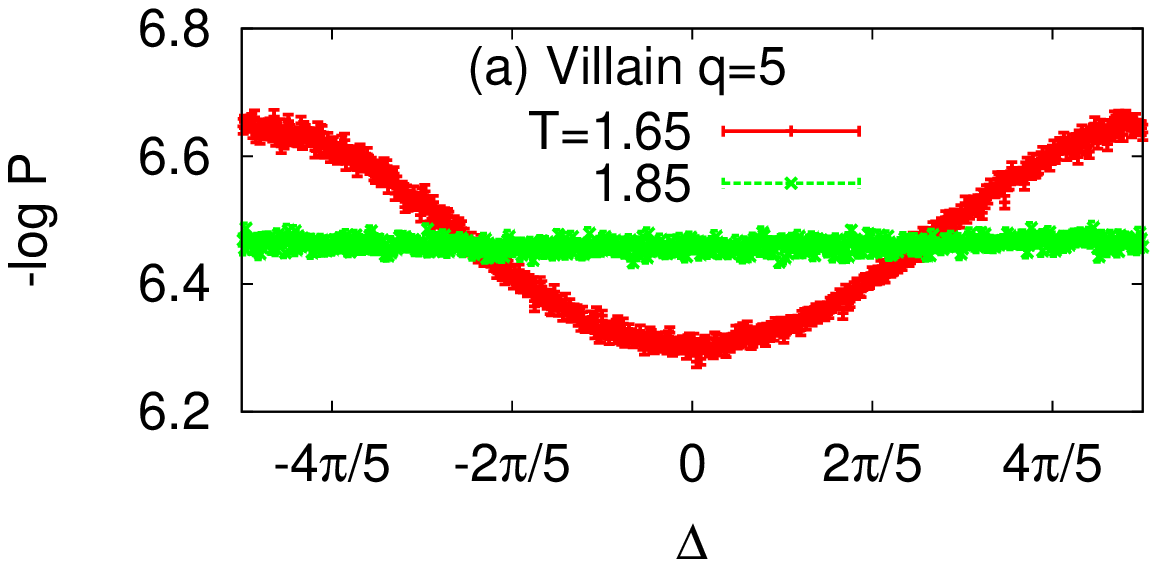}
\includegraphics[width=0.45\textwidth]{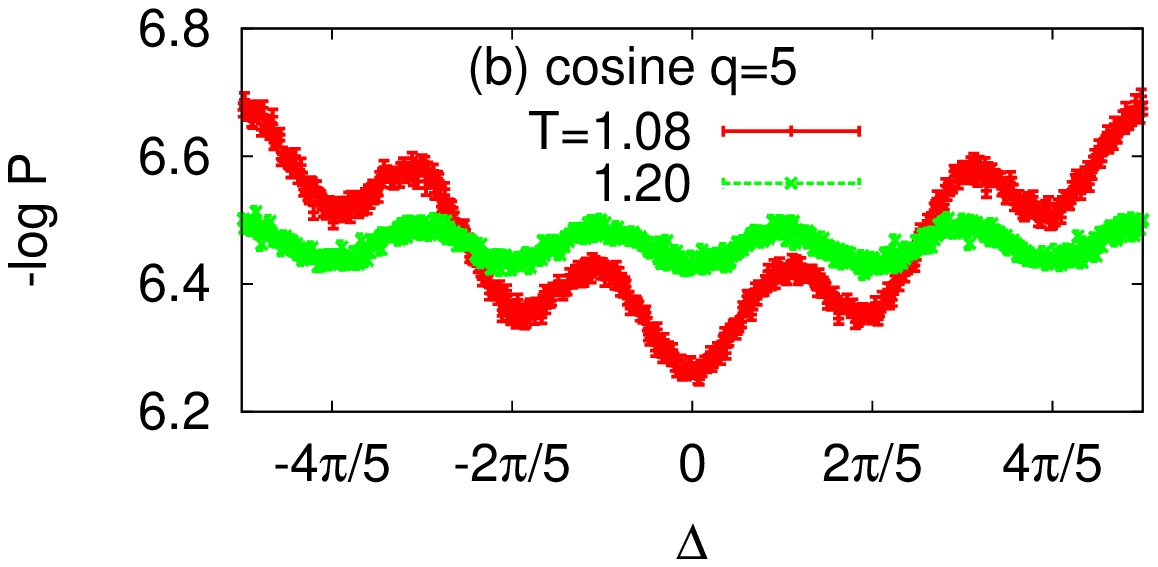}
\includegraphics[width=0.45\textwidth]{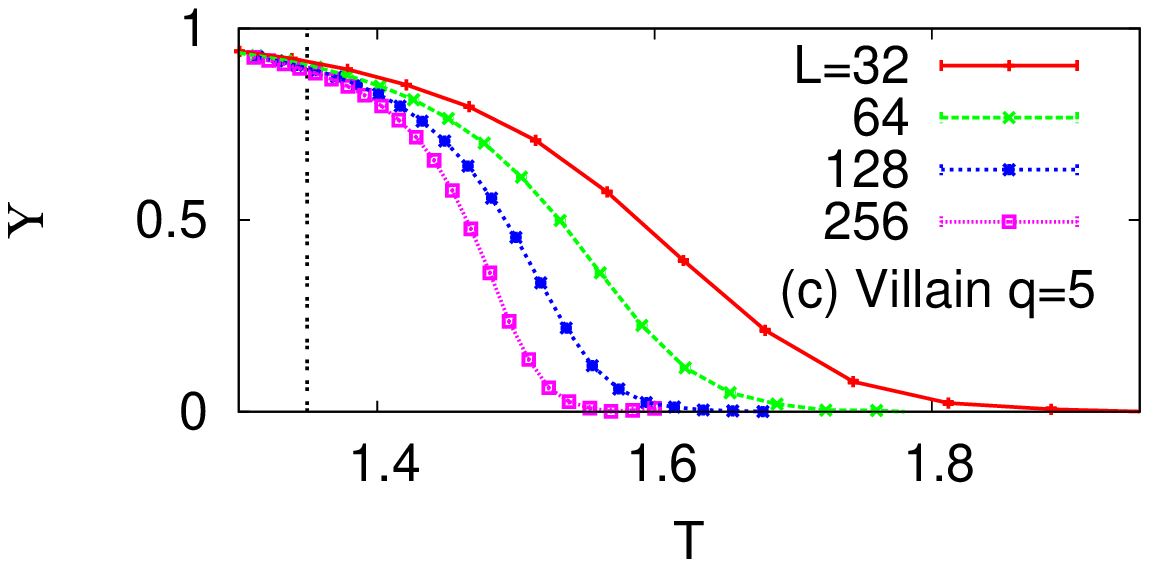}
\includegraphics[width=0.45\textwidth]{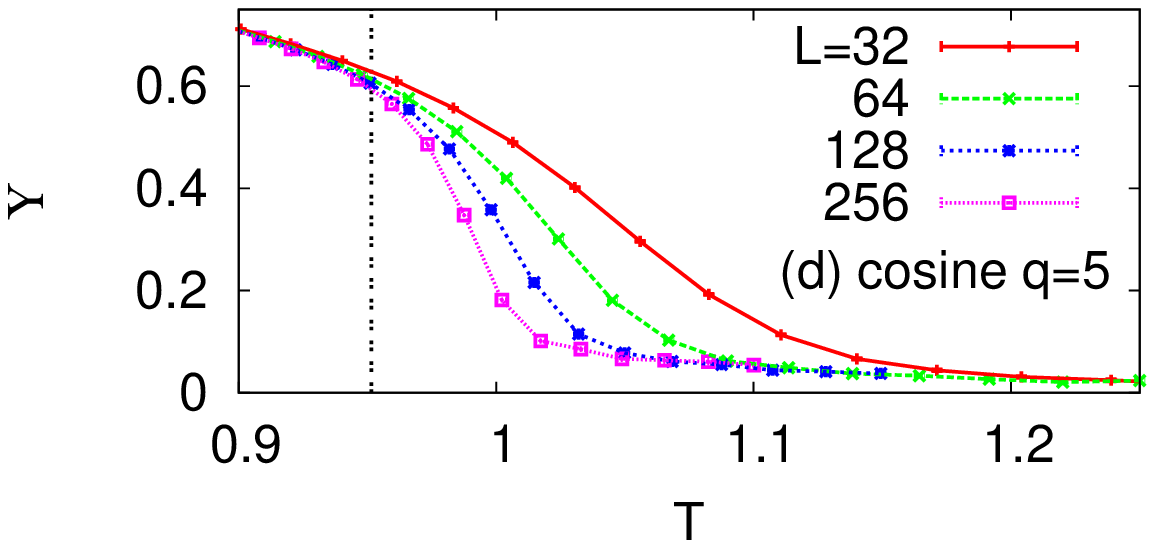}
\caption{(Color online) (a) Free energy $F(\Delta)$ of the Villain
five-state clock model [cf. Fig.~\ref{fig:ftbc}(c)]. (b) Five-fold symmetry
is clearly visible in the five-state clock model with the cosine potential.
We take $L=32$ as the system size in both the cases. (c) As the system size
grows in the Villain five-state clock model, the change in curvature at
$\Delta=0$ [i.e., Eq.~(\ref{eq:y})] becomes sharper, approaching the universal
jump to zero. The vertical line shows the critical temperature $T_{c2}^{\rm
Villain} \approx 1.35$ estimated by extrapolating the size dependence of the
universal jump condition $\Upsilon(T_c^{\rm Villain}) = 2T_c^{\rm
Villain}/\pi$~\cite{nelson,*minnhagen} according to
the KT scenario $\tau \sim (\ln L)^{-2}$.
(d) When $L$ increases in the five-state clock model, the
curvature at $\Delta=0$ remains finite even above $T_{c2} \approx
0.9515$~\cite{papa1}, which is represented as the vertical line.}
\label{fig:five}
\end{figure}

In the six-state clock model, on the other hand, we expect from the
vanishing $\Upsilon$ that the
six-fold symmetry will be manifested only at $T < T_{c1}$ and is not visible at
higher temperatures. In the language of the RG,
it means that the system can respond to perturbations as
if it were a continuous spin system on average over large length scales.
Our numerical calculations on the $L\times L$ square lattice clearly support
this picture as depicted in Figs.~\ref{fig:ftbc}(c) and \ref{fig:ftbc}(d).
Note that these results are obtained with quite a small size, i.e., $L=32$, so
that the transition temperature $T_{c2}$ tends to be overestimated
[see, e.g., Figs.~\ref{fig:five}(c) or \ref{fig:five}(d)]. Nevertheless,
these give us a
qualitatively correct picture that remains true for larger system sizes (see
Ref.~\cite{clock}, which shows data up to $L=512$). In short, the six-state
clock model effectively exhibits the continuous $U(1)$ symmetry whereby the
genuine KT behavior becomes possible.

It is interesting to see that the Villain five-state clock model
also shows the same kind of behavior, supporting theoretical works
based on the Villain formulation [Fig.~\ref{fig:five}(a)]. However, when
the method is applied to the conventional five-state clock model to obtain
$F(\Delta)$, we see that the behavior is more similar to what we observed
for $q=2$ or $q=4$ than to $q=6$, in that the model exhibits its five-fold
symmetry even in the disordered phase [Fig.~\ref{fig:five}(b)]. Note that
the free energy $F(\Delta)$ has a curvature near $\Delta = 0$, which
corresponds to the nonvanishing $\Upsilon$. Therefore, the nonvanishing
$\Upsilon$ signals the residual discrete symmetry above the upper transition,
which remains there even if the system size grows [compare
Fig.~\ref{fig:five}(c) and \ref{fig:five}(d)],
so we can
conclude that the upper phase transition of the five-state clock model is
not of the conventional KT type.

\begin{figure}
\includegraphics[width=0.45\textwidth]{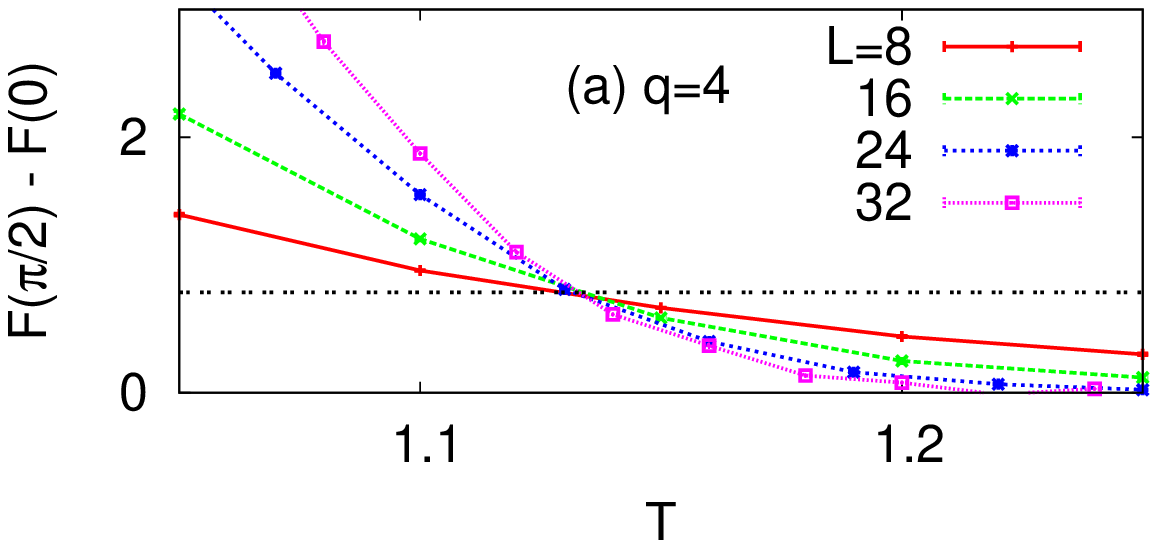}
\includegraphics[width=0.45\textwidth]{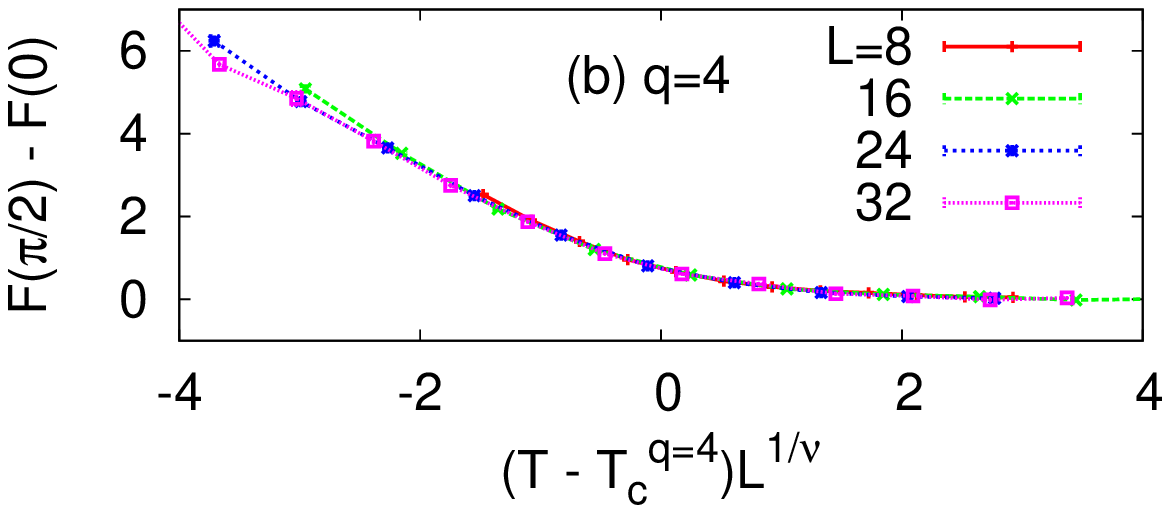}
\includegraphics[width=0.45\textwidth]{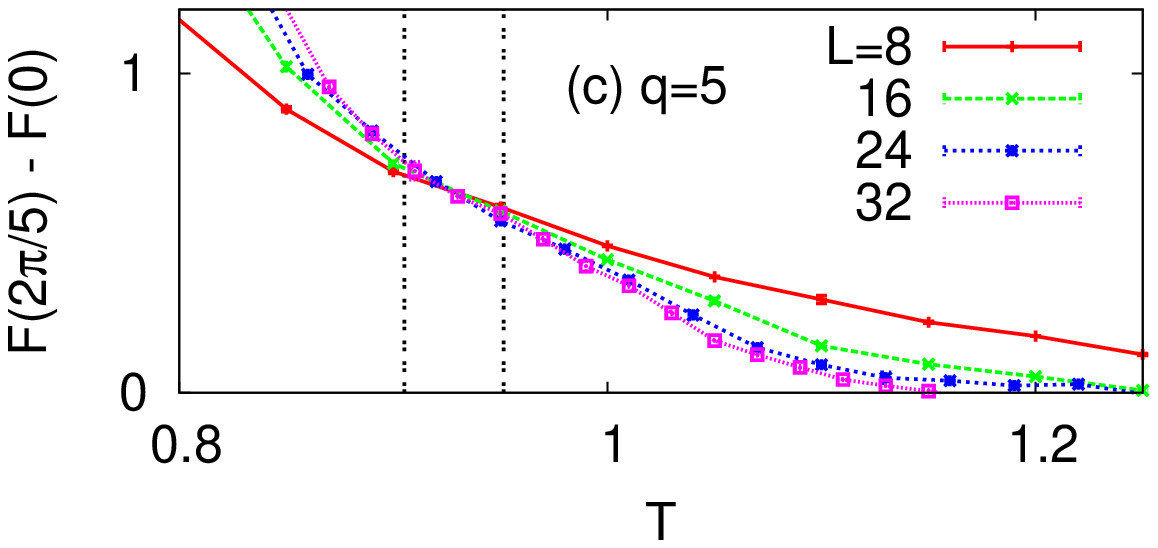}
\includegraphics[width=0.45\textwidth]{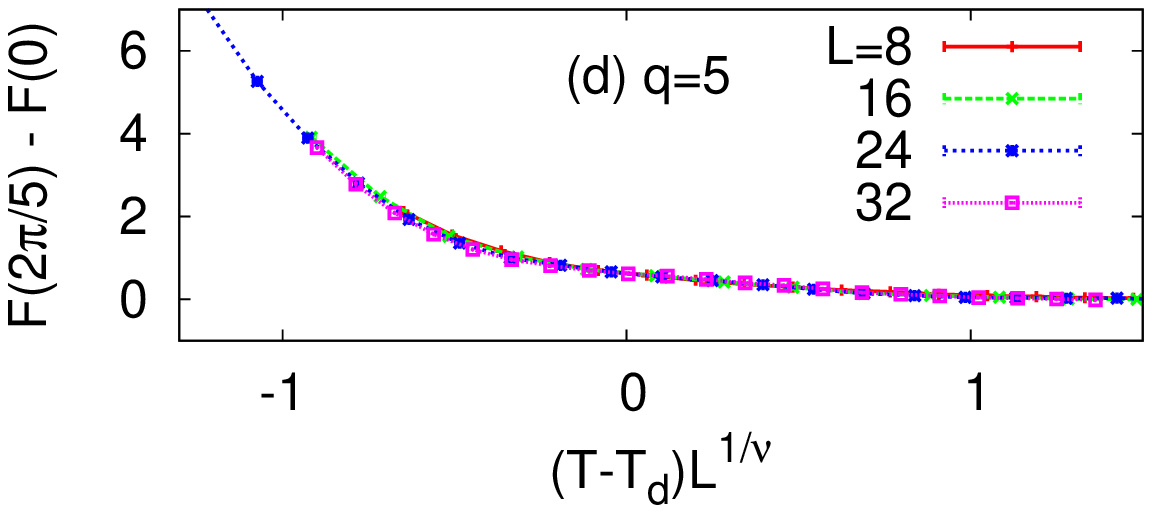}
\caption{(Color online) (a) Interfacial free energy of the four-state clock
model as a function of $T$ and (b) its scaling collapse where the exact values
$T_c^{q=4} = 1/\ln(1+\sqrt{2}) \approx 1.135$ and $\nu=1$ are used. The
horizontal line represents $\delta F = \pi/4$. (c)
Interfacial free energy of the five-state clock model, where the two
vertical lines represent $T_{c1}$ and $T_{c2}$ measured in
Ref.~\cite{papa1}, respectively. (d) Scaling collapse attempted with the
dual point $T_d \approx 0.9291$~\cite{nish} and a trial exponent $\nu =
2$ used in Ref.~\cite{papa1}. Error bars are usually smaller than the symbol
sizes.}
\label{fig:interface}
\end{figure}

As a brief sideline, we may consider what one would find if $\Delta$ also
had the discrete $q$-fold symmetry.
For the Ising model, this free-energy difference between the periodic
boundary condition ($\Delta=0$) and the anti-periodic boundary condition
($\Delta = \pi$) is called the interfacial free energy.
This quantity is known to have the following scaling form~\cite{night}:
\begin{equation}
\delta F = f \left(\tau L^{1/\nu} \right),
\label{eq:interface}
\end{equation}
where $f$ is a scaling function with a universal amplitude $f(0) = \pi/4$
and $\nu$ is the scaling
exponent for the divergence of the correlation length $\xi \sim \tau^{-\nu}$.
It is straightforward to generalize the notion of the interfacial free
energy to the case of the $q$-state clock model by measuring the free-energy
difference caused by setting $\Delta = 2\pi/q$. In other words, we define
$\delta F \equiv F(2\pi/q) - F(0)$, and this quantity shows how
the lower envelope of $F(\Delta)$ changes its shape. Note that 
the envelope is not distinguished from $F(\Delta)$ if $q \ge 6$ so that we
do not have to consider such cases separately.
For the four-state clock model, we observe that this method
indeed yields correct scaling results [Figs.~\ref{fig:interface}(a) and
\ref{fig:interface}(b)]. When applied to the five-state clock model, $\delta
F$ does vanish in the high-temperature phase by construction in contrast to
the helicity modulus [Fig.~\ref{fig:interface}(c)]. The question is then how
this quantity describes the critical properties. In Ref.~\cite{papa1},
the authors attempt scaling collapse with $\nu = 2$ under the assumption
that the model undergoes a usual continuous phase transition. Using this
trial exponent, we also observe reasonable scaling collapse around the dual
temperature $T_d \approx 0.9291$~\cite{nish} as depicted in
Fig.~\ref{fig:interface}(d), but the possibility of a crossover phenomenon
cannot be ruled out for such small sizes.

\section{Discussion and Summary}
\label{sec:discuss}

\begin{figure}
\includegraphics[width=0.45\textwidth]{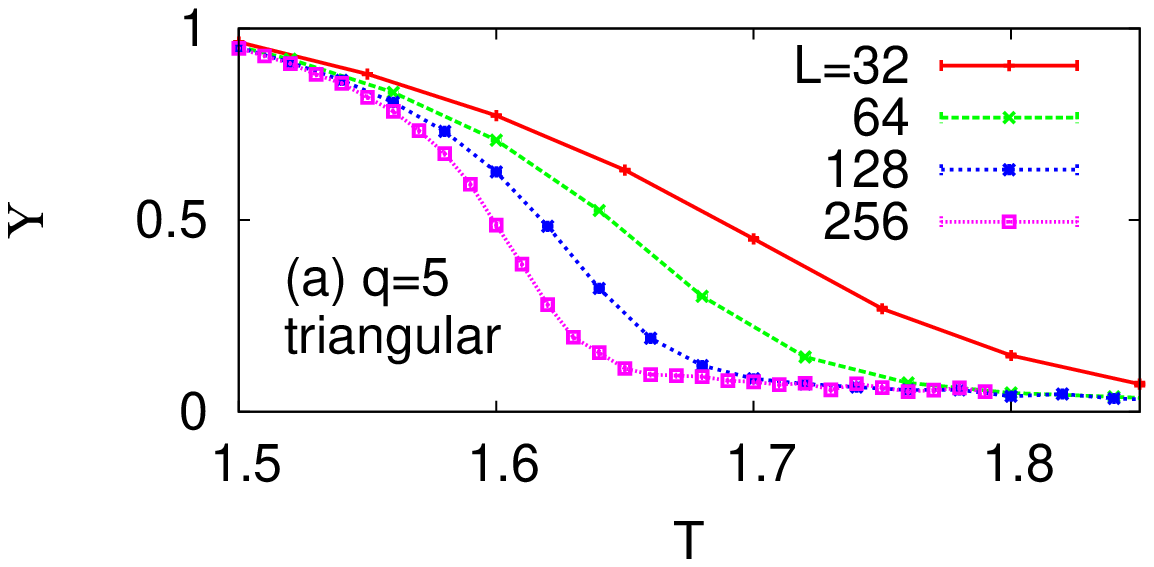}
\includegraphics[width=0.45\textwidth]{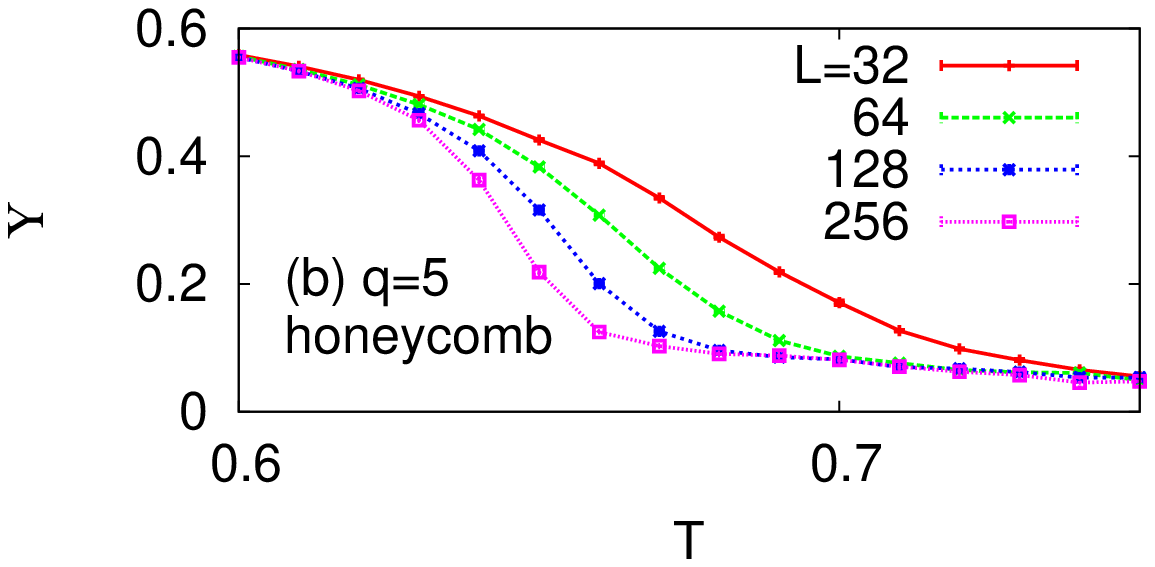}
\caption{(Color online) Helicity modulus [Eq.~(\ref{eq:y})] of the
five-state clock model (a) on the triangular lattice and (b) on the
honeycomb lattice. Error bars are not larger than the symbol sizes. See also
Ref.~\cite{stroud} about computing $\Upsilon$ on these lattices.}
\label{fig:q5}
\end{figure}

A possible reason for the peculiarity of the five-state clock model
has already been argued in
Refs.~\cite{marcel1,marcel2}. This theory points out a possibility that the
model may have two different length scales, one from individual vortices and
the other from composites of vortices. According to this scenario, the
nonvanishing helicity modulus is a finite-size effect governed by the larger
length scale of the composites, which will eventually become negligible in the
thermodynamic limit. The correlation due to the composites is destroyed
at a higher temperature called the disorder line, roughly estimated as
$T_{\rm disorder} \approx [1-\cos(4\pi/5)] \approx 1.81$, where the
correlated liquid becomes a Potts gas.
It turns out hard to examine the existence of the
second length scale with numerical calculations:
If there really was a second length scale, it would have to get
shorter with increasing $T$ so that the helicity modulus would then be more
sensitive to the lattice size. For this reason, we have tried to observe the
helicity modulus as a function of $L$ with fixing the temperature at $T=1.1$
well above $T_{c2} \approx 0.9515$. However, it appears to converge to a
nonzero value at least up to $L = 1024$ (see also Ref.~\cite{papa2}),
implying that the second
length scale, if it exists, might be still greater than this size at such
high $T$. We additionally note that
Ref.~\cite{marcel2} suggests a \emph{conspiracy} between the clock number
$q=5$ and the four-fold symmetry of the square lattice:
For a unit-charge vortex located at a square plaquette,
the sum of angle differences around it is $2\pi$,
and there are only \emph{four} angle differences, each of which can take an
integer multiple of $2\pi/5$. It means that there should be at least one
angle difference greater than $2\pi/5$, likely to be
$4\pi/5$. So the argument for the origin of the composite on the square
lattice is that such a high-energy
domain wall tends to appear as a double strand of two single domain walls in
the vicinity of the vortex-antivortex unbinding transition, forming a
correlated local structure (see Ref.~\cite{domany,*baltar,*knops} for
related discussions). However, it is not easy to extend such an argument to
other 2D lattices as already noted in Ref.~\cite{marcel1}, and it turns out
that we need a more universal explanation because
the nonvanishing helicity modulus is observed on the triangular and the
honeycomb lattices as well (Fig.~\ref{fig:q5}).
Therefore, our observation is only partially consistent with the
theory in Refs.~\cite{marcel1,marcel2} at the moment, and it seems quite
nontrivial to explain how the discrete five-fold symmetry emerges in
vortex composites over more than $10^3$ lattice spacings.
If we once again borrow the hand-waving RG argument in
Sec.~\ref{sec:intro}~\cite{einhorn}, our observation rather implies another
possibility that the discreteness in the spin degree of freedom is not
completely renormalized away with the cosine potential and that this may
survive further RG transformations.

In summary, we have found the origin of the nonvanishing helicity modulus
numerically
observed in the five-state clock model: It originates from the five-fold
symmetry, which is not replaced by the continuous $U(1)$ symmetry even deep
inside the high-temperature phase and clearly manifests itself under the FTBC.
As a consequence, although our transfer-matrix calculation supports
the three-phase picture for the five-state clock model,
the nature of the transition is not fully developed to the genuine KT type,
differently from its Villain-approximated version, in our Monte Carlo
calculations.
Since there is not enough numerical evidence for a vanishing helicity
modulus in the five-state clock model, it is only for $q \ge 6$ that one can
be sure of the three phases separated by a genuine KT transition and its
dual, and we suggest that the transition of the five-state clock model is
a weaker cousin of the KT transition.

\acknowledgments
S.K.B. is thankful to H. Park for bringing his attention to
Ref.~\cite{marcel1}. Conversations with M. den Nijs and J. M. Kosterlitz are
also gratefully
acknowledged. H.M. was supported by the Academy of Finland through 
its Centres of Excellence Program (Project No. 251748). B.J.K. was supported by the National Research
Foundation of Korea (NRF) grant funded by the Korean government (MEST) (Grant
No. 2011-0015731). This work was supported by the Supercomputing
Center/Korea Institute of Science and Technology Information with
supercomputing resources including technical support (Project No.
KSC-2012-C1-05).

%
\end{document}